\documentclass[12pt,preprint,showpacs,preprintnumbers]{revtex4}
\usepackage{amssymb}
\usepackage{amsmath}
\usepackage{graphicx}
\usepackage{dcolumn}
\usepackage{bm}
\usepackage{epsfig}
\usepackage[T1]{fontenc}
\usepackage{ae,aecompl}

\begin{document}

\title{Two dimensional MRT LB model for compressible and incompressible flows}
\author{Feng Chen$^1$\footnote{
Corresponding author. E-mail: shanshiwycf@163.com}, Aiguo
Xu$^2$\footnote{ Corresponding author. E-mail:
Xu\_Aiguo@iapcm.ac.cn}, Guangcai Zhang$^2$, Yonglong Wang$^1$}
\affiliation{1,School of science, Institute of condensed matter
physics, Linyi University, Linyi 276005, China \\
2, National Key Laboratory of Computational Physics, Institute of Applied
Physics and Computational Mathematics, P. O. Box 8009-26, Beijing 100088,
P.R.China}
\date{\today }

\begin{abstract}
In the paper we extend the Multiple-Relaxation-Time (MRT) Lattice
Boltzmann (LB) model proposed in [Europhys. Lett. \textbf{90},
54003 (2010)] so that it is suitable also for incompressible
flows. To decrease the artificial oscillations, the convection
term is discretized by the flux limiter scheme with splitting
technique. New model is validated by some well-known benchmark
tests, including Riemann problem and Couette flow, and satisfying
agreements are obtained between the simulation results and
analytical ones. In order to show the merit of LB model over
traditional methods, the non-equilibrium characteristics of system
are solved. The simulation results are consistent with physical
analysis.
\end{abstract}

 \maketitle

\section{Introduction}

In recent years, the Lattice Boltzmann (LB) method has emerged as an
attractive computational approach for complex physical system\cite%
{SS,BSV,XGL1}. The lattice Bhatnagar-Gross-Krook (BGK) model,
based on a single-relaxation-time approximation, is the simplest
and the most popular form. However, this simplicity also leads to
some deficiencies, such as the numerical stability problem, and
fixed Prandtl number. To overcome these deficiencies of BGK model,
the Multiple-Relaxation-Time (MRT) lattice Boltzmann
method\cite{HSB,13} has been developed, and successfully used in
simulating various fluid flow
problems\cite%
{Lallemand1,Lallemand2,PRE036701,JCP539,CF957,CMA1392,JCP453,JCP7774,PRE016705,JPA055501}%
. Most of the existing MRT models work only for isothermal system.
To simulate system with temperature field, many attempts have been
made\cite{pre036706,pre026705,Mezrhab2010}.

Besides the models mentioned above, we proposed a MRT Finite
Difference lattice Boltzmann model for compressible flows with
arbitrary specific heat ratio and Prandtl number in previous
work\cite{chenepl}. In the model, the kinetic moment space and the
equilibria of nonconserved moments are constructed according to
the seven-moment relations associated with the local equilibrium
distribution function. Numerical experiments showed that
compressible flows with strong shocks can be well simulated by
this model.

In the paper we extend the MRT LB model so that it is suitable
also for incompressible flows. In order to efficiently decrease
the unphysical oscillations, the flux limiter
scheme\cite{Sofonea1,Sofonea3,chenctp} with splitting technique is
incorporated into the new model. When the system deviates more
from equilibrium, the LB simulation can give more physical
information\cite{xufop,yan,JCP}, such as the non-equilibrium
characteristics of system. Here, in the new MRT LB model, the
non-equilibrium characteristics of system are solved through a
dynamic procedure where a shock wave propagates from a heavy
medium to a light one.

The rest of the paper is organized as follows. Section II presents
the extended MRT LB model. Section III describes the finite
difference schemes. section IV is for the validation and
verification of the new LB model. Non-equilibrium characteristics
are shown and analyzed in section V. Section VI makes the
conclusion for the present paper.

\section{Model description}

According to the main strategy of MRT LB method, the MRT LB equation can be
described as:
\begin{equation}
\frac{\partial f_{i}}{\partial t}+v_{i\alpha }\frac{\partial f_{i}}{\partial
x_{\alpha }}=-\mathbf{M}_{il}^{-1}\hat{\mathbf{S}}_{lk}(\hat{f}_{k}-\hat{f}%
_{k}^{eq})\text{,}  \label{3}
\end{equation}%
where $f_{i}$ and $\hat{f}_{i}$ are the particle distribution function in
the velocity space and the kinetic moment space respectively, $\mathbf{v}%
_{i} $ is the discrete particle velocity, $i=1$,$\ldots$ ,$N$, $N$\ is the
number of discrete velocities, the subscript $\alpha $\ indicates $x$\ or $y$%
. The matrix $\hat{\mathbf{S}}=\mathbf{MSM}^{-1}=diag(s_{1},s_{2},\cdots
,s_{N})$ is the diagonal relaxation matrix. $\mathbf{M}$ is the
transformation matrix between the velocity space and the kinetic moment
space. $\hat{f}_{i}=m_{ij}f_{j}$, $m_{ij}$ is an element of the
transformation matrix. $\hat{f}_{i}^{eq}$ is the equilibrium value of
distribution function $\hat{f}_{i}$ in the kinetic moment space.

\begin{figure}[tbp]
\center\includegraphics*[width=0.40\textwidth]{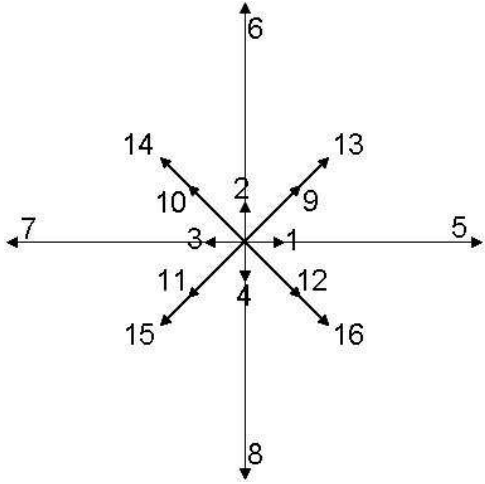}
\caption{ Schematics of $\mathbf{v}_{i}$ for the discrete velocity model.}
\end{figure}

In the previous work, we constructed a two-dimensional MRT LB model based on
the model by Kataoka and Tsutahara\cite{Kataoka} (see Fig. 1):
\begin{equation*}
\left( v_{i1,}v_{i2}\right) =\left\{
\begin{array}{cc}
\mathbf{cyc}:\left( \pm 1,0\right) , & \text{for }1\leq i\leq 4, \\
\mathbf{cyc}:\left( \pm 6,0\right) , & \text{for }5\leq i\leq 8, \\
\sqrt{2}\left( \pm 1,\pm 1\right) , & \text{for }9\leq i\leq 12, \\
\frac{3}{\sqrt{2}}\left( \pm 1,\pm 1\right) , & \text{for }13\leq i\leq 16,%
\end{array}%
\right.
\end{equation*}%
where \textbf{cyc} indicates the cyclic permutation. Transformation matrix $%
\mathbf{M}$ and the equilibrium distribution function $\hat{f}_{i}^{eq}$ in
the moment space are chosen according to the seven-moment relations (see
Appendix for details). At the continuous limit, the above formulation
recovers the following Navier-Stokes (NS) equations:
\begin{subequations}
\begin{equation}
\frac{\partial \rho }{\partial t}+\frac{\partial (\rho u_{x})}{\partial x}+%
\frac{\partial (\rho u_{y})}{\partial y}=0,
\end{equation}%
\begin{eqnarray}
&&\frac{\partial (\rho u_{x})}{\partial t}+\frac{\partial }{\partial x}(\rho
u_{x}^{2})+\frac{\partial }{\partial y}(\rho u_{x}u_{y})  \notag \\
&=&-\frac{\partial P}{\partial x}+\frac{\partial }{\partial y}[\frac{\rho RT%
}{s_{7}}(\frac{\partial u_{y}}{\partial x}+\frac{\partial u_{x}}{\partial y}%
)]  \notag \\
&&+\frac{\partial }{\partial x}[\frac{\rho RT}{s_{5}}(1-\frac{2}{b})(\frac{%
\partial u_{x}}{\partial x}+\frac{\partial u_{y}}{\partial y})+\frac{\rho RT%
}{s_{6}}(\frac{\partial u_{x}}{\partial x}-\frac{\partial u_{y}}{\partial y}%
)]\text{,}
\end{eqnarray}%
\begin{eqnarray}
&&\frac{\partial (\rho u_{y})}{\partial t}+\frac{\partial }{\partial x}(\rho
u_{x}u_{y})+\frac{\partial }{\partial y}(\rho u_{y}^{2})  \notag \\
&=&-\frac{\partial P}{\partial y}+\frac{\partial }{\partial x}[\frac{\rho RT%
}{s_{7}}(\frac{\partial u_{y}}{\partial x}+\frac{\partial u_{x}}{\partial y}%
)]  \notag \\
&&+\frac{\partial }{\partial y}[\frac{\rho RT}{s_{5}}(1-\frac{2}{b})(\frac{%
\partial u_{x}}{\partial x}+\frac{\partial u_{y}}{\partial y})-\frac{\rho RT%
}{s_{6}}(\frac{\partial u_{x}}{\partial x}-\frac{\partial u_{y}}{\partial y}%
)]\text{,}
\end{eqnarray}%
\begin{eqnarray}
&&\frac{\partial e}{\partial t}+\frac{\partial }{\partial x}[(e+2P)u_{x}]+%
\frac{\partial }{\partial y}[(e+2P)u_{y}]  \notag \\
&=&2\frac{\partial }{\partial x}\{\frac{\rho RT}{s_{8}}[(\frac{b}{2}+1)R%
\frac{\partial T}{\partial x}+(2\frac{\partial u_{x}}{\partial x}-\frac{2}{b}%
\frac{\partial u_{x}}{\partial x}-\frac{2}{b}\frac{\partial u_{y}}{\partial y%
})u_{x}+(\frac{\partial u_{y}}{\partial x}+\frac{\partial u_{x}}{\partial y}%
)u_{y}]\}  \notag \\
&&+2\frac{\partial }{\partial y}\{\frac{\rho RT}{s_{9}}[(\frac{b}{2}+1)R%
\frac{\partial T}{\partial y}+(\frac{\partial u_{y}}{\partial x}+\frac{%
\partial u_{x}}{\partial y})u_{x}+(2\frac{\partial u_{y}}{\partial y}-\frac{2%
}{b}\frac{\partial u_{x}}{\partial x}-\frac{2}{b}\frac{\partial u_{y}}{%
\partial y})u_{y}]\}\text{.}
\end{eqnarray}%
where $P=\rho RT$, $e=b\rho RT+\rho u_{\alpha }^{2}$ is twice of
the total energy, and $b$\ is a constant related to the
specific-heat-ratio $\gamma =(b+2)/b$.

In order to maintain the isotropy constraint of viscous stress tensor and
heat conductivity, some of the relaxation parameters should be equal to one
another, namely $s_{5}=s_{6}=s_{7}$, $s_{8}=s_{9}$. The above NS equations
reduce to
\end{subequations}
\begin{subequations}
\begin{equation}
\frac{\partial \rho }{\partial t}+\frac{\partial (\rho u_{\alpha })}{%
\partial x_{\alpha }}=0\text{,}
\end{equation}%
\begin{equation}
\frac{\partial (\rho u_{\alpha })}{\partial t}+\frac{\partial \left( \rho
u_{\alpha }u_{\beta }\right) }{\partial x_{\beta }}=-\frac{\partial P}{%
\partial x_{\alpha }}+\frac{\partial }{\partial x_{\beta }}[(\mu \frac{%
\partial u_{\alpha }}{\partial x_{\beta }}+\frac{\partial u_{\beta }}{%
\partial x_{\alpha }}-\frac{2}{3}\frac{\partial u_{\chi }}{\partial x_{\chi }%
}\delta _{\alpha \beta })+\mu _{B}\frac{\partial u_{\chi }}{\partial x_{\chi
}}\delta _{\alpha \beta }]\text{,}  \label{3b}
\end{equation}%
\begin{equation}
\frac{\partial e}{\partial t}+\frac{\partial }{\partial x_{\alpha }}\left[
(e+2P)u_{\alpha }\right] =2\frac{\partial }{\partial x_{\beta }}[(\frac{b}{2}%
+1)\lambda^{\prime } R\frac{\partial T}{\partial x_{\beta }}+\lambda^{\prime
} (\frac{\partial u_{\alpha }}{\partial x_{\beta }}+\frac{\partial u_{\beta }%
}{\partial x_{\alpha }}-\frac{2}{b}\frac{\partial u_{\chi }}{\partial
x_{\chi }}\delta _{\alpha \beta })u_{\alpha }]\text{,}  \label{3c}
\end{equation}
where the viscosity $\mu =\rho RT/s_{5}$, the bulk viscosity $\mu
_{B}=(2/3-2/b) \rho RT/s_{5}$, $\lambda^{\prime } =\rho RT/s_{8}$, $\left(
\alpha, \beta, \chi=x,y\right)$.

However, the viscous coefficient in the energy equation \eqref{3c}
is not consistent with that in the momentum equation \eqref{3b}.
By modifying the collision operators of the moments related to
energy flux:
\end{subequations}
\begin{subequations}
\begin{align}
& \hat{\mathbf{S}}_{88}(\hat{f}_{8}-\hat{f}_{8}^{eq})\Rightarrow \hat{%
\mathbf{S}}_{88}(\hat{f}_{8}-\hat{f}_{8}^{eq})+(s_{8}/s_{5}-1)\rho Tu_{x}
\notag \\
& \qquad \times (4\frac{\partial u_{x}}{\partial x}-\frac{4}{b}\frac{%
\partial u_{x}}{\partial x}-\frac{4}{b}\frac{\partial u_{y}}{\partial y}%
)+(s_{8}/s_{5}-1)\rho Tu_{y}(2\frac{\partial u_{y}}{\partial x}+2\frac{%
\partial u_{x}}{\partial y})\text{,}
\end{align}%
\begin{align}
& \hat{\mathbf{S}}_{99}(\hat{f}_{9}-\hat{f}_{9}^{eq})\Rightarrow \hat{%
\mathbf{S}}_{99}(\hat{f}_{9}-\hat{f}_{9}^{eq})+(s_{9}/s_{5}-1)\rho Tu_{x}
\notag \\
& \qquad \times (2\frac{\partial u_{y}}{\partial x}+2\frac{\partial u_{x}}{%
\partial y})+(s_{9}/s_{5}-1)\rho Tu_{y}(4\frac{\partial u_{y}}{\partial y}-%
\frac{4}{b}\frac{\partial u_{x}}{\partial x}-\frac{4}{b}\frac{\partial u_{y}%
}{\partial y})\text{,}
\end{align}%
we get the following energy equation:
\end{subequations}
\begin{equation}
\frac{\partial e}{\partial t}+\frac{\partial }{\partial x_{\alpha }}\left[
(e+2P)u_{\alpha }\right] =2\frac{\partial }{\partial x_{\beta }}[\lambda
\frac{\partial T}{\partial x_{\beta }}+\mu (\frac{\partial u_{\alpha }}{%
\partial x_{\beta }}+\frac{\partial u_{\beta }}{\partial x_{\alpha }}-\frac{2%
}{b}\frac{\partial u_{\chi }}{\partial x_{\chi }}\delta _{\alpha \beta
})u_{\alpha }]\text{,}
\end{equation}%
where the thermal conductivity $\lambda =(\frac{b}{2}+1)R\lambda^{\prime }$.

\section{Finite Difference Scheme}

In the original LB model\cite{chenepl}, the time evolution is
based on the usual first-order forward Euler scheme, while space
discretization is performed through a Lax-Wendroff scheme. In this
work, the flux limiter scheme with splitting technique
corresponding to the MRT model is adopted. The proposed flux
limiter scheme can efficiently decrease the unphysical
oscillations around the interfaces.

\begin{figure}[tbp]
\center\includegraphics*[width=0.65\textwidth]{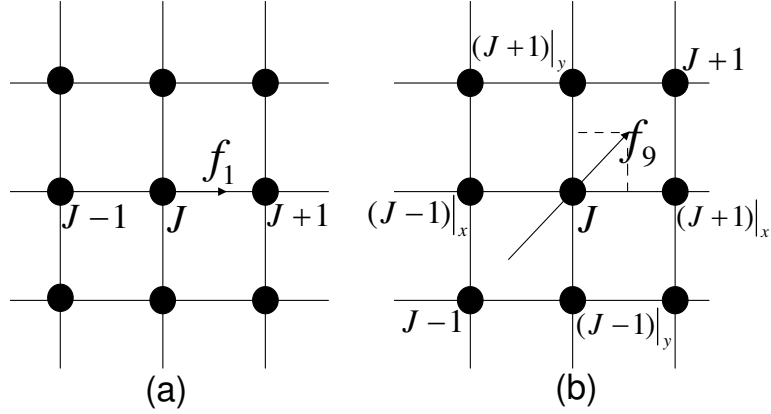}
\caption{Characteristic lines and corresponding projections in the
$x$ and $y $ directions. (a): $f_{1}(\mathbf{x},t)$; (b):
$f_{9}(\mathbf{x},t)$.}
\end{figure}
Figure 2 shows the characteristic lines in the flux limiter scheme and
corresponding projections in $x$ and $y$ directions. $%
\left( J-1)\right\vert _{x}$ and $\left( J-1)\right\vert _{y}$ are
corresponding projections of node $J-1$ in the $x$ and $y$ directions. Let $%
f_{i,J}^{n}$ be the value of distribution function at time $t$ in
the node $J$ along the direction $i$, we rewrite the evolution of
$f_{i}$ in node $J$ at time step $t+dt$ as follows,
\begin{equation}
f_{i,J}^{n+1}=f_{i,J}^{n}-\frac{dt}{A_{i}dx}[\left.
F_{i,J+1/2}^{n}\right\vert _{x}-\left. F_{i,J-1/2}^{n}\right\vert _{x}]-%
\frac{dt}{A_{i}dy}[\left. F_{i,J+1/2}^{n}\right\vert _{y}-\left.
F_{i,J-1/2}^{n}\right\vert _{y}]-dt\mathbf{M}_{il}^{-1}\hat{\mathbf{S}}_{lk}(%
\hat{f}_{k,J}^{n}-\hat{f}_{k,J}^{n,eq})\text{,}
\end{equation}%
where
\begin{equation}
A_{i}=\left\{
\begin{array}{cc}
1, & \text{for }1\leq i\leq 4, \\
1/6, & \text{for }5\leq i\leq 8, \\
1/\sqrt{2}, & \text{for }9\leq i\leq 12, \\
\sqrt{2}/3, & \text{for }13\leq i\leq 16.%
\end{array}%
\right.
\end{equation}%
$\left. F_{i,J+1/2}^{n}\right\vert _{x}$ ($\left. F_{i,J-1/2}^{n}\right\vert
_{x}$) and $\left. F_{i,J+1/2}^{n}\right\vert _{y}$ ($\left.
F_{i,J-1/2}^{n}\right\vert _{y}$) are $x$ and $y$ components of the outgoing
(incoming) flux in node $J$ along the direction $i$,
\begin{subequations}
\begin{equation}
\left. F_{i,J+1/2}^{n}\right\vert _{x}=f_{i}^{n}(ix,iy)+\frac{1}{2}(1-\frac{%
dt}{A_{i}dx})[f_{i}^{n}(ix+A_{i}v_{ix},iy)-f_{i}^{n}(ix,iy)]\psi _{x}(ix,iy)%
\text{,}
\end{equation}%
\begin{equation}
\left. F_{i,J-1/2}^{n}\right\vert _{x}=f_{i}^{n}(ix-A_{i}v_{ix},iy)+\frac{1}{%
2}(1-\frac{dt}{A_{i}dx})[f_{i}^{n}(ix,iy)-f_{i}^{n}(ix-A_{i}v_{ix},iy)]\psi
_{x}(ix-A_{i}v_{ix},iy)\text{,}
\end{equation}%
\begin{equation}
\left. F_{i,J+1/2}^{n}\right\vert _{y}=f_{i}^{n}(ix,iy)+\frac{1}{2}(1-\frac{%
dt}{A_{i}dy})[f_{i}^{n}(ix,iy+A_{i}v_{iy})-f_{i}^{n}(ix,iy)]\psi _{y}(ix,iy)%
\text{,}
\end{equation}%
\begin{equation}
\left. F_{i,J-1/2}^{n}\right\vert _{y}=f_{i}^{n}(ix,iy-A_{i}v_{iy})+\frac{1}{%
2}(1-\frac{dt}{A_{i}dy})[f_{i}^{n}(ix,iy)-f_{i}^{n}(ix,iy-A_{i}v_{iy})]\psi
_{y}(ix,iy-A_{i}v_{iy})\text{.}
\end{equation}%
The flux limiter is expressed as
\end{subequations}
\begin{equation}
\psi _{\alpha }(ix,iy)=\left\{
\begin{array}{ccc}
0 & \text{,} & \left. \theta _{i}^{n}(ix,iy)\right\vert _{\alpha }\leq 0 \\
2\left. \theta _{i}^{n}(ix,iy)\right\vert _{\alpha } & \text{,} & 0\leq
\left. \theta _{i}^{n}(ix,iy)\right\vert _{\alpha }\leq \frac{1}{3} \\
(1+\left. \theta _{i}^{n}(ix,iy)\right\vert _{\alpha })/2 & \text{,} & \frac{%
1}{3}\leq \left. \theta _{i}^{n}(ix,iy)\right\vert _{\alpha }\leq 3 \\
2 & \text{,} & 3\leq \left. \theta _{i}^{n}(ix,iy)\right\vert _{\alpha }%
\end{array}%
\right.
\end{equation}%
where the smoothness functions are
\begin{subequations}
\begin{equation}
\left. \theta _{i}^{n}(ix,iy)\right\vert _{x}=\frac{%
f_{i}^{n}(ix,iy)-f_{i}^{n}(ix-A_{i}v_{ix},iy)}{%
f_{i}^{n}(ix+A_{i}v_{ix},iy)-f_{i}^{n}(ix,iy)}\text{,}
\end{equation}%
\begin{equation}
\left. \theta _{i}^{n}(ix,iy)\right\vert _{y}=\frac{%
f_{i}^{n}(ix,iy)-f_{i}^{n}(ix,iy-A_{i}v_{iy})}{%
f_{i}^{n}(ix,iy+A_{i}v_{iy})-f_{i}^{n}(ix,iy)}\text{.}
\end{equation}%
\end{subequations}
The Lax-Wendroff scheme is recovered for the flux limiter $\psi
_{x}=\psi _{y}=1$, and the first order upwind scheme is recovered
when $\psi _{x}=\psi _{y}=0$.

\section{validation and verification}

\subsection{Performance on discontinuity}

\begin{figure}[tbp]
\center\includegraphics*[bbllx=15pt,bblly=15pt,bburx=315pt,bbury=230pt,width=0.65\textwidth]{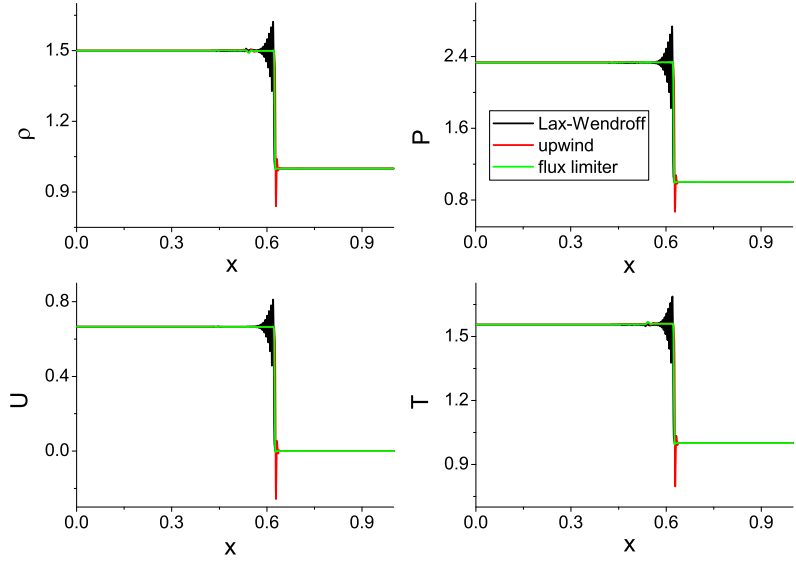}
\caption{Simulation results with various difference schemes at
$t=0.06$.}
\end{figure}
In order to check the performance of flux limiter scheme on
discontinuity, we construct the following problem
\begin{equation}
\left\{
\begin{array}{cc}
(\rho ,u_{1},u_{2},T)=(1.5,0.666667,0.0,1.55556), & x\leq L/2. \\
(\rho ,u_{1},u_{2},T)=(1.0,0.0,0.0,1.0), & L/2<x<L.%
\end{array}%
\right.
\end{equation}
$L$ is the length of computational domain. In the $x$ direction,
$f_{i}=\mathbf{M}_{ij}^{-1}\hat{f}_{j}^{eq}$ is set, where the
macroscopic quantities adopt the initial values. In the $y$
direction, the periodic boundary condition is adopted. The
physical quantities on the two sides satisfy the Hugoniot
relations. Fig. 3 shows the simulation results of density,
pressure, $x-$ component of velocity, and temperature at time
$t=0.06$ using different space discretization schemes. The
parameters are $\gamma=2$, $dx=dy=0.001$, $dt=10^{-5}$,
$s_{5}=s_{6}=s_{7}=5\times 10^{4}$, and other collision parameters
are $10^{5}$. The simulations with Lax-Wendroff scheme have strong
unphysical oscillations in the shocked region. The second order
upwind scheme results in unphysical `overshoot' phenomena at the
shock front. The simulation results with flux limiter scheme are
much more accurate, and this scheme has the ability to decrease
the unphysical oscillations at the discontinuity.

\subsection{Lax shock tube problem}

\begin{figure}[tbp]
\center\includegraphics*[bbllx=15pt,bblly=15pt,bburx=320pt,bbury=230pt,width=0.65\textwidth]{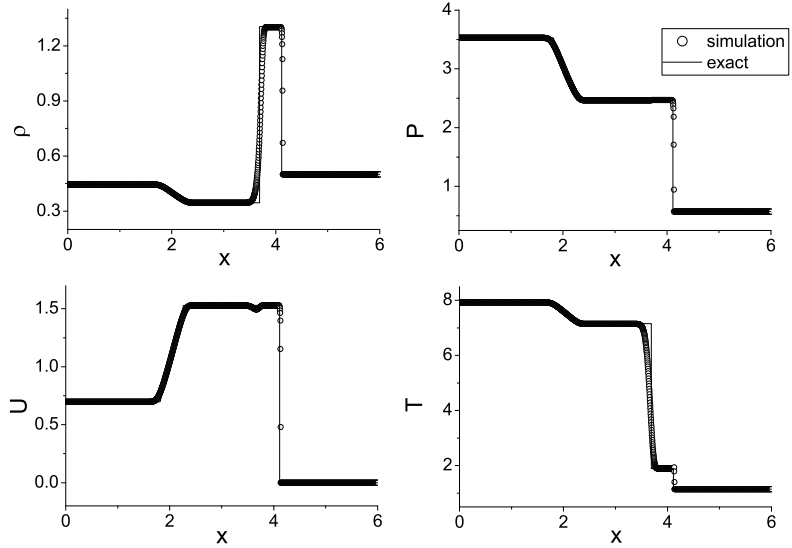}
\caption{LB simulation results and exact solutions for Lax shock
tube at $t=0.45$.}
\end{figure}
The initial condition of the problem is:
\begin{equation}
\left\{
\begin{array}{cc}
(\rho ,u_{1},u_{2},T)=(0.445,0.698,0.0,7.928), & x\leq L/2. \\
(\rho ,u_{1},u_{2},T)=(0.5,0.0,0.0,1.142), & L/2<x<L.%
\end{array}%
\right.
\end{equation}
The profiles of density, pressure, $x-$ component of velocity, and
temperature at $t=0.45$ are shown in Fig. 4, where the exact
solutions are presented with solid lines for comparison. The
parameters are
$\gamma =1.4$, $dx=dy=0.003$, $dt=10^{-5}$, $s_{5}=s_{6}=s_{7}=2\times 10^{3}$, $%
s_{8}=s_{9}=10^{3}$, and other collision parameters are $10^{5}$.
Obviously, the simulation results agree well with the exact
solutions.

The above simulations show that compressible flows, especially
those with discontinuity and shock waves, can be well simulated by
the present model.

\subsection{Couette flow}

\begin{figure}[tbp]
\begin{center}
\includegraphics[bbllx=10pt,bblly=15pt,bburx=340pt,bbury=190pt,
scale=0.9]{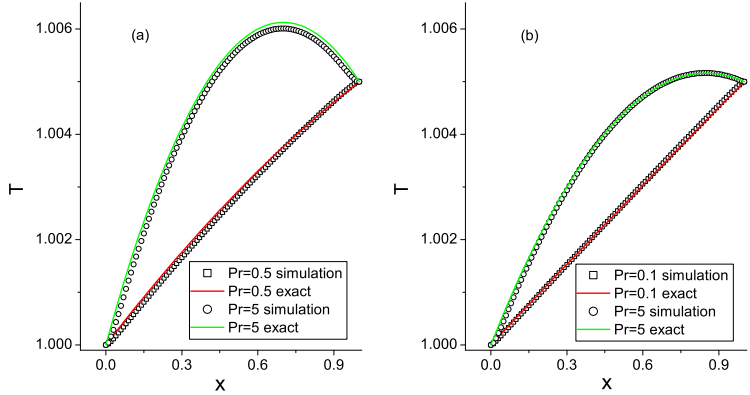}
\end{center}
\caption{Temperature profiles of Couette flow. (a) $\gamma=2$,
$Pr=0.5$ corresponds to $s_{5}=10^3,s_{8}=5\times10^2$, $Pr=5$
corresponds to $s_{5}=2\times10^2,s_{8}=10^3$, and (b)
$\gamma=1.4$, $Pr=0.1$ corresponds to $s_{5}=10^3,s_{8}=10^2$,
$Pr=5$ corresponds to $s_{5}=2\times10^2,s_{8}=10^3$ (other
collision parameters are $10^3$).}
\end{figure}
Here we conduct a series of numerical simulations of Couette flow.
In the simulation, the left wall is fixed and the right wall moves
at speed $u_{x}=0$, $u_{y}=0.1$. The initial state of the fluid is
$\rho =1$, $T=1$, $u_{x}=0$, $u_{y}=0$. The simulation results are
compared with the analytical solution:
\begin{align}
T=T_{1}+(T_{2}-T_{1})\frac{x}{H}+\frac{\mu }{2\lambda }u_{y}^{2}\frac{x}{H}(1-%
\frac{x}{H})\text{,}
\end{align}
where $T_{1}$ and $T_{2}$ are temperatures of the left and right walls ($%
T_{1}=1$, $T_{2}=1.005$), $H$ is the width of the channel.
Periodic boundary conditions are applied to the bottom and top
boundaries, and the left and right walls adopt the nonequilibrium
extrapolation method. Fig. 5 shows the comparison of LB results
with analytical solutions for thermal Couette Flows. (a)
corresponds to $\gamma=2$, and (b) corresponds to $\gamma=1.4$. It
is clearly shown that the simulation results of new model are in
agreement with the analytical solutions, and the Prandtl number
effects are successfully captured. New model is suitable for
incompressible flows.

\section{Non-equilibrium characteristic}

To show the merit of LB method over traditional ones, in this
section we study the non-equilibrium characteristics using the new
model. Among the moment relations required by each LB model, only
for the first three (density, momentum and energy), the equilibrium distribution function $%
f_{i}^{eq}$ can be replaced by the distribution function $f_{i}$.
If we replace $f_{i}^{eq}$ by $f_{i}$ in the left hand of other
moment relations, the value of left side will have a difference
from that of the right side. This difference represents the
deviation of system from its thermodynamic equilibrium\cite{xufop,yan,JCP}.
In this MRT LB model,
the kinetic moment space and the corresponding equilibria of
nonconserved moments are constructed according to the seven-moment
relations. So, the deviation from equilibrium in this model can be
defined as $\Delta
_{i}=\hat{f}_{i}-\hat{f}_{i}^{eq}=\textbf{M}_{ij}(f_{j}-f_{j}^{eq})$.
$\Delta _{i}$ contains the information of macroscopic flow
velocity $u_{\alpha}$. Furthermore, we replace $v_{i\alpha}$ by
$v_{i\alpha}-u_{\alpha}$ in the transformation matrix
$\textbf{M}$, named $\textbf{M}^{\ast}$ (see Appendix for
details). $\Delta
_{i}^{\ast}=\textbf{M}_{ij}^{\ast}(f_{j}-f_{j}^{eq})$ is only the
manifestation of molecular thermalmotion and does not contain the
information of macroscopic flow.

\begin{figure}[tbp]
\center\includegraphics*[bbllx=10pt,bblly=15pt,bburx=325pt,bbury=230pt,width=0.9\textwidth]{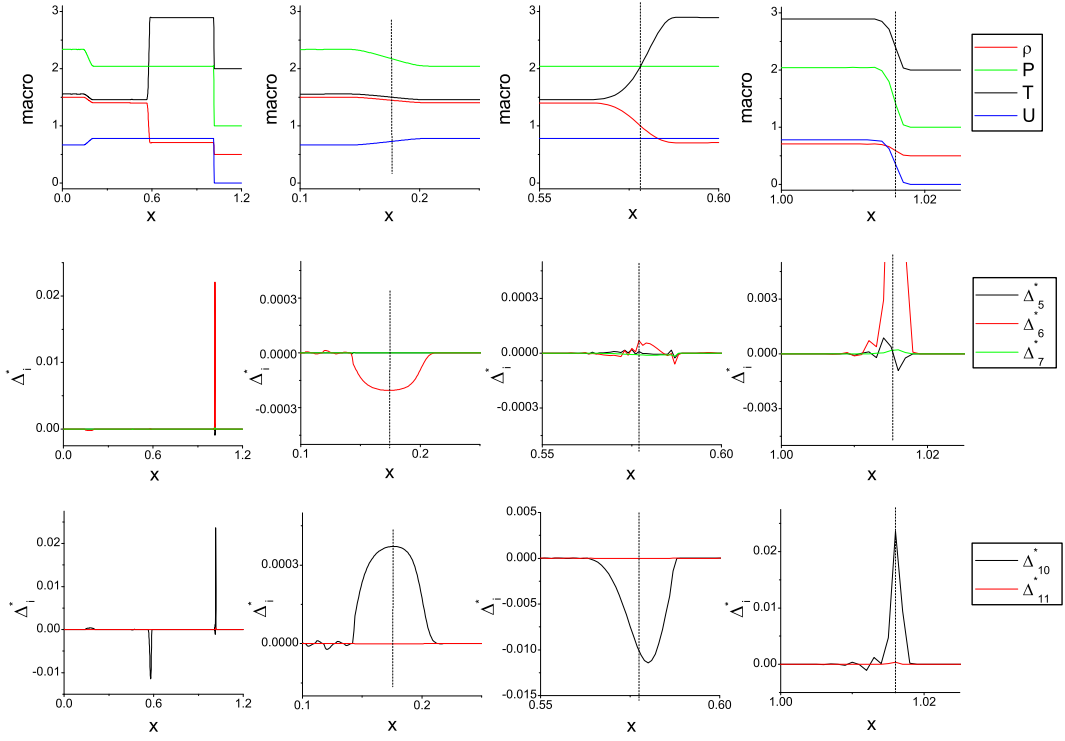}
\caption{LB numerical results and non-equilibrium characteristics
at $t=0.3$.}
\end{figure}
Now, we study the following dynamic procedure. An incident shock
wave with Mach number $1.414$ travels from a heavy medium and hits
a light one, where the two different fluids are separated by an
unperturbed interface. The initial macroscopic quantities are as
follows:
\begin{equation*}
\left\{
\begin{array}{cc}
(\rho,u_{1},u_{2},p)_{s}=(1.5,0.666667,0,2.33334) \text{,} &  \\
(\rho,u_{1},u_{2},p)_{h}=(1,0,0,1) \text{,} &  \\
(\rho,u_{1},u_{2},p)_{l}=(0.5,0,0,1) \text{,} &
\end{array}%
\right.
\end{equation*}%
where the subscripts $s$, $h$, $l$ indicate the shock wave region, the heavy
medium region, and the light medium region. In our simulations, the
computational domain is $[0, 1.2]\times[0, 0.01]$, and divided into $%
1200\times10$ mesh-cells. The initial position of shock wave is $x=0.24$,
the unperturbed interface lies at the position $x=0.4$. Inflow boundary is
applied at the left side, outflow boundary is applied at the right side, and
periodic boundary conditions are applied at the top and bottom boundaries. $%
\gamma =2$ in the whole domain. The density, pressure, $x-$
component of velocity and temperature profiles and $\Delta
_{i}^{\ast}$ ($i=5,6,7,10,11$) on the center line $y=0.005$ at
time $t=0.3$ are shown in Fig. 6. The parameters are $dt=10^{-5}$,
$s_{5}=s_{6}=s_{7}=5\times 10^{4}$, and other collision parameters
are $10^{5}$.

In the figures, the system shows three different interfaces,
rarefaction wave, material interface, and shock wave. Physical
quantities change significantly at the three interfaces, and
vertical lines indicate the positions of interfaces. The system
starts to deviate from equilibrium once the physical quantities
starts to change. When the physical quantities arrives at its
steady-state required by the Hugoniot relations, the system goes
back to its equilibrium state. The peak values of deviations
$\Delta _{i}^{\ast}$ at shock wave interface are larger than the
others. This is because the shock dynamic procedure is faster than
the other two processes, and the system has less time to relax to
its thermodynamic equilibrium.

At the interfaces, $\Delta _{5}^{\ast}$, $\Delta _{7}^{\ast}$ and
$\Delta _{11}^{\ast}$ have small amplitudes. $\Delta _{5}^{\ast}$
contains two parts, $x$ and $y$ components of internal translational kinetic
energy. This indicates that the two parts deviate from equilibrium
in opposite directions with the same amplitude. $\Delta
_{6}^{\ast}$ shows an opposite deviation for the rarefaction wave
interface and the shock interface. The physical reason is as
below. The temperature gradient first initiates variance of the
internal kinetic energy in the direction of temperature gradient.
(Here, the temperature shows gradient in the $x$ direction.) Then,
part of internal kinetic energy variance is transferred to other
degrees of freedoms via collisions of molecules. The internal
kinetic energy in the temperature gradient direction further
varies, and so on. The shock wave increases density, pressure and
temperature, while the rarefaction wave decreases those
quantities. So, $\Delta _{6}^{\ast}$ shows a negative deviation
for the rarefaction wave interface, while shows a positive
deviation for the shock interface. The values of $\Delta
_{10}^{\ast}$ at material interface and shock wave interface have
the same order, and are much larger than that at rarefaction wave.
This is because the sizes of temperature variation near the
material interface and shock wave differ little, and larger than
that near the rarefaction wave. When the temperature gradient
vanishes, the system attains its thermodynamic equilibrium.

\section{Conclusions}

In the paper a MRT LB model for compressible flows is extended so
that it is suitable also for incompressible flows. In order to
efficiently decrease the unphysical oscillations, space
discretization adopts flux limiter scheme with splitting
technique. It is validated and verified via same well-known
benchmark tests, including Riemann problem and Couette flow, and
satisfying agreements are obtained between the new model results
and analytical ones. In order to show the merit of LB model over
traditional methods, we studied the behaviors of system deviating
from its equilibrium through a dynamic procedure where shock wave
propagates from a heavy material to a light one. The simulation
results are consistent with the physical analysis.

\section*{Acknowledgements}

The authors would like to sincerely thank S. Succi and C. Lin for many instructive
discussions. We acknowledge support of National Natural Science Foundation
of China[under Grant Nos. 11075021 and 11047020]. AX and GZ acknowledge
support of the Science Foundation of CAEP [under Grant Nos. 2012B0101014 and
2011A0201002] and the Foundation of State Key Laboratory of Explosion
Science and Technology[under Grant No.KFJJ14-1M].

\begin{appendix}

\section{Transformation matrix and equilibria of the nonconserved moments}

In the model by Kataoka and Tsutahara, the local equilibrium
distribution function $f_{i}^{eq}$ satisfies the following
relations:
\begin{subequations}
\begin{equation}
\rho =\sum f_{i}^{eq},
\end{equation}%
\begin{equation}
\rho u_{\alpha }=\sum f_{i}^{eq}v_{i\alpha },
\end{equation}%
\begin{equation}
\rho \left( bRT+u_{\alpha }^{2}\right) =\sum f_{i}^{eq}\left(
v_{i\alpha }^{2}+\eta _{i}^{2}\right) ,
\end{equation}%
\begin{equation}
P\delta _{\alpha \beta }+\rho u_{\alpha }u_{\beta }=\sum
f_{i}^{eq}v_{i\alpha }v_{i\beta },
\end{equation}%
\begin{equation}
\rho \left[ \left( b+2\right) RT+u_{\beta }^{2}\right] u_{\alpha
}=\sum f_{i}^{eq}\left( v_{i\beta }^{2}+\eta _{i}^{2}\right)
v_{i\alpha },
\end{equation}%
\begin{equation}
\rho \left[ RT\left( u_{\alpha }\delta _{\beta \chi }+u_{\beta
}\delta _{\alpha \chi }+u_{\chi }\delta _{\alpha \beta }\right)
+u_{\alpha }u_{\beta }u_{\chi }\right] =\sum f_{i}^{eq}v_{i\alpha
}v_{i\beta }v_{i\chi },
\end{equation}%
\begin{equation}
\rho \left\{ \left( b+2\right) R^{2}T^{2}\delta _{\alpha \beta
}+\left[
\left( b+4\right) u_{\alpha }u_{\beta }+u_{\chi }^{2}\delta _{\alpha \beta }%
\right] RT+u_{\chi }^{2}u_{\alpha }u_{\beta }\right\} =\sum
f_{i}^{eq}\left( v_{i\chi }^{2}+\eta _{i}^{2}\right) v_{i\alpha
}v_{i\beta },
\end{equation}%
\end{subequations}
where a parameter $\eta _{i}$ is introduced, in order to describe
the $(b-2)$ extra-degrees of freedom corresponding to molecular
rotation and/or vibration, where $\eta _{i}=5/2$ for $i=1$,
$\cdots $,$4$, and $\eta _{i}=0$ for $i=5$, $\cdots $, $16$.

The transformation matrix $\mathbf{M}$ in the MRT model is
composed as below: $\mathbf{M}=(m_{1},m_{2},\cdots ,m_{16})^{T}$,
\begin{subequations}
\begin{equation}
m_{1i}=1\text{,}
\end{equation}%
\begin{equation}
m_{2i}=v_{ix}\text{,}
\end{equation}%
\begin{equation}
m_{3i}=v_{iy}\text{,}
\end{equation}%
\begin{equation}
m_{4i}=v_{ix}^{2}+v_{iy}^{2}+\eta _{i}^{2}\text{,}
\end{equation}%
\begin{equation}
m_{5i}=v_{ix}^{2}+v_{iy}^{2}\text{,}
\end{equation}%
\begin{equation}
m_{6i}=v_{ix}^{2}-v_{iy}^{2}\text{,}
\end{equation}%
\begin{equation}
m_{7i}=v_{ix}v_{iy}\text{,}
\end{equation}%
\begin{equation}
m_{8i}=v_{ix}(v_{ix}^{2}+v_{iy}^{2}+\eta _{i}^{2})\text{,}
\end{equation}%
\begin{equation}
m_{9i}=v_{iy}(v_{ix}^{2}+v_{iy}^{2}+\eta _{i}^{2})\text{,}
\end{equation}%
\begin{equation}
m_{10i}=v_{ix}(v_{ix}^{2}+v_{iy}^{2})\text{,}
\end{equation}%
\begin{equation}
m_{11i}=v_{iy}(v_{ix}^{2}+v_{iy}^{2})\text{,}
\end{equation}%
\begin{equation}
m_{12i}=v_{ix}(v_{ix}^{2}-v_{iy}^{2})\text{,}
\end{equation}%
\begin{equation}
m_{13i}=v_{iy}(v_{ix}^{2}-v_{iy}^{2})\text{,}
\end{equation}%
\begin{equation}
m_{14i}=(v_{ix}^{2}+v_{iy}^{2})(v_{ix}^{2}+v_{iy}^{2}+\eta
_{i}^{2})\text{,}
\end{equation}%
\begin{equation}
m_{15i}=v_{ix}v_{iy}(v_{ix}^{2}+v_{iy}^{2}+\eta _{i}^{2})\text{,}
\end{equation}%
\begin{equation}
m_{16i}=(v_{ix}^{2}-v_{iy}^{2})(v_{ix}^{2}+v_{iy}^{2}+\eta
_{i}^{2})\text{,}
\end{equation}%
\end{subequations}
where $i=1,\cdots ,16$.

Replacing $v_{i\alpha}$ by $v_{i\alpha}-u_{\alpha}$ in the
transformation matrix $\textbf{M}$, matrix $\mathbf{M}^{\ast}$ is
expressed as follows:
$\mathbf{M}^{\ast}=(m_{1}^{\ast},m_{2}^{\ast},\cdots
,m_{16}^{\ast})^{T}$,
\begin{subequations}
\begin{equation}
m_{1i}^{\ast}=1\text{,}
\end{equation}%
\begin{equation}
m_{2i}^{\ast}=v_{ix}-u_{x}\text{,}
\end{equation}%
\begin{equation}
m_{3i}^{\ast}=v_{iy}-u_{y}\text{,}
\end{equation}%
\begin{equation}
m_{4i}^{\ast}=(v_{ix}-u_{x})^{2}+(v_{iy}-u_{y})^{2}+\eta
_{i}^{2}\text{,}
\end{equation}%
\begin{equation}
m_{5i}^{\ast}=(v_{ix}-u_{x})^{2}+(v_{iy}-u_{y})^{2}\text{,}
\end{equation}%
\begin{equation}
m_{6i}^{\ast}=(v_{ix}-u_{x})^{2}-(v_{iy}-u_{y})^{2}\text{,}
\end{equation}%
\begin{equation}
m_{7i}^{\ast}=(v_{ix}-u_{x})(v_{iy}-u_{y})\text{,}
\end{equation}%
\begin{equation}
m_{8i}^{\ast}=(v_{ix}-u_{x})[(v_{ix}-u_{x})^{2}+(v_{iy}-u_{y})^{2}+\eta
_{i}^{2}]\text{,}
\end{equation}%
\begin{equation}
m_{9i}^{\ast}=(v_{iy}-u_{y})[(v_{ix}-u_{x})^{2}+(v_{iy}-u_{y})^{2}+\eta
_{i}^{2}]\text{,}
\end{equation}%
\begin{equation}
m_{10i}^{\ast}=(v_{ix}-u_{x})[(v_{ix}-u_{x})^{2}+(v_{iy}-u_{y})^{2}]\text{,}
\end{equation}%
\begin{equation}
m_{11i}^{\ast}=(v_{iy}-u_{y})[(v_{ix}-u_{x})^{2}+(v_{iy}-u_{y})^{2}])\text{,}
\end{equation}%
\begin{equation}
m_{12i}^{\ast}=(v_{ix}-u_{x})[(v_{ix}-u_{x})^{2}-(v_{iy}-u_{y})^{2}]\text{,}
\end{equation}%
\begin{equation}
m_{13i}^{\ast}=(v_{iy}-u_{y})[(v_{ix}-u_{x})^{2}-(v_{iy}-u_{y})^{2}]\text{,}
\end{equation}%
\begin{equation}
m_{14i}^{\ast}=[(v_{ix}-u_{x})^{2}+(v_{iy}-u_{y})^{2}][(v_{ix}-u_{x})^{2}+(v_{iy}-u_{y})^{2}+\eta
_{i}^{2}]\text{,}
\end{equation}%
\begin{equation}
m_{15i}^{\ast}=(v_{ix}-u_{x})(v_{iy}-u_{y})[(v_{ix}-u_{x})^{2}+(v_{iy}-u_{y})^{2}+\eta
_{i}^{2}]\text{,}
\end{equation}%
\begin{equation}
m_{16i}^{\ast}=[(v_{ix}-u_{x})^{2}-(v_{iy}-u_{y})^{2}][(v_{ix}-u_{x})^{2}+(v_{iy}-u_{y})^{2}+\eta
_{i}^{2}]\text{,}
\end{equation}%
\end{subequations}
where $i=1,\cdots ,16$.

The equilibria of nonconserved moments are as follows:
\begin{subequations}
\begin{equation}
\hat{f}_{5}^{eq}=2P+(j_{x}^{2}+j_{y}^{2})/\rho \text{,}
\end{equation}%
\begin{equation}
\hat{f}_{6}^{eq}=(j_{x}^{2}-j_{y}^{2})/\rho \text{,}
\end{equation}%
\begin{equation}
\hat{f}_{7}^{eq}=j_{x}j_{y}/\rho \text{,}
\end{equation}%
\begin{equation}
\hat{f}_{8}^{eq}=(e+2P)j_{x}/\rho \text{,}
\end{equation}%
\begin{equation}
\hat{f}_{9}^{eq}=(e+2P)j_{y}/\rho \text{,}
\end{equation}%
\begin{equation}
\hat{f}_{10}^{eq}=(4P+j_{x}^{2}/\rho +j_{y}^{2}/\rho )j_{x}/\rho
\text{,}
\end{equation}%
\begin{equation}
\hat{f}_{11}^{eq}=(4P+j_{x}^{2}/\rho +j_{y}^{2}/\rho )j_{y}/\rho
\text{,}
\end{equation}%
\begin{equation}
\hat{f}_{12}^{eq}=(2P+j_{x}^{2}/\rho -j_{y}^{2}/\rho )j_{x}/\rho
\text{,}
\end{equation}%
\begin{equation}
\hat{f}_{13}^{eq}=(-2P+j_{x}^{2}/\rho -j_{y}^{2}/\rho )j_{y}/\rho
\text{,}
\end{equation}%
\begin{equation}
\hat{f}_{14}^{eq}=2(b+2)\rho
R^{2}T^{2}+(6+b)RT(j_{x}^{2}+j_{y}^{2})/\rho
+(j_{x}^{2}+j_{y}^{2})^{2}/\rho ^{3}\text{,}
\end{equation}%
\begin{equation}
\hat{f}_{15}^{eq}=[(b+4)P+(j_{x}^{2}+j_{y}^{2})/\rho ]j_{x}j_{y}/\rho ^{2}%
\text{,}
\end{equation}%
\begin{equation}
\hat{f}_{16}^{eq}=[(b+4)P+(j_{x}^{2}+j_{y}^{2})/\rho
](j_{x}^{2}-j_{y}^{2})/\rho ^{2}\text{.}
\end{equation}%
\end{subequations}

\end{appendix}


\end{document}